\documentclass[aps,prl,twocolumn,notitlepage,showpacs,superscriptaddress,10pt]{revtex4-1}%
\usepackage{graphicx}
\usepackage{amsmath}
\usepackage{amssymb}
\usepackage{color}
\usepackage{amsfonts}%
\usepackage[caption=false]{subfig}

\usepackage{color}
\usepackage[dvipsnames]{xcolor}
\usepackage[colorlinks,bookmarks=false,citecolor=darkblue,linkcolor=red,urlcolor=blue]{hyperref}
\usepackage[capitalize]{cleveref}

\definecolor{darkred}{rgb}{0.7,0.0,0.0}

\definecolor{darkblue}{rgb}{0,0.02,0.45}

\definecolor{darkgreen}{rgb}{0.02,0.45,0.0}

\definecolor{violet}{rgb}{0.8,0.2,0.6}

\providecommand{\U}[1]{\protect\rule{.1in}{.1in}}
\newcommand{\nairo}{Na$_2$IrO$_3$}
\newcommand{\liiro}{$\alpha$-Li$_2$IrO$_3$}
\newcommand{\hliiro}{H$_3$LiIr$_2$O$_6$}
\newcommand{\rucl}{$\alpha$-RuCl$_3$}

\begin{document}

\title{Modified Curie-Weiss Law for $j_{\rm eff}$ Magnets}

\author{Ying Li}\thanks{yingli1227@xjtu.edu.cn}
\affiliation{MOE Key Laboratory for Nonequilibrium Synthesis and Modulation of Condensed Matter, School of Physics, Xi'an Jiaotong University, Xi'an 710049, China}
\affiliation{Institut f\"ur Theoretische Physik, Goethe-Universit\"at Frankfurt,
Max-von-Laue-Strasse 1, 60438 Frankfurt am Main, Germany}
\author{Stephen M. Winter}\thanks{winters@wfu.edu}
\affiliation{Department of Physics and Center for Functional Materials, Wake Forest University, NC 27109, USA}
\author{David A. S. Kaib}
\affiliation{Institut f\"ur Theoretische Physik, Goethe-Universit\"at Frankfurt,
Max-von-Laue-Strasse 1, 60438 Frankfurt am Main, Germany}
\author{Kira Riedl}
\affiliation{Institut f\"ur Theoretische Physik, Goethe-Universit\"at Frankfurt,
Max-von-Laue-Strasse 1, 60438 Frankfurt am Main, Germany}
\author{Roser Valent{\'\i}}\thanks{valenti@itp.uni-frankfurt.de}
\affiliation{Institut f\"ur Theoretische Physik, Goethe-Universit\"at Frankfurt,
Max-von-Laue-Strasse 1, 60438 Frankfurt am Main, Germany}

\date{\today}

\begin{abstract}
In spin-orbit-coupled magnetic materials, the usually applied Curie-Weiss law can break down. This is due to potentially sharp temperature-dependence of the local magnetic moments. 
We therefore propose a modified Curie-Weiss formula suitable for analysis of experimental susceptibility. We show for octahedrally coordinated materials of $d^5$ filling 
that the Weiss constant obtained from the improved formula is in excellent agreement with the calculated Weiss constant from microscopic exchange interactions. Reanalyzing the measured susceptibility of several Kitaev candidate materials with the modified formula resolves apparent discrepancies between various experiments regarding the magnitude and anisotropies of the underlying magnetic couplings. 
\end{abstract}
\maketitle
\par

Great interest has been devoted towards searching for Kitaev spin liquid candidate materials with strongly anisotropic Ising couplings on the honeycomb lattice
~\cite{kitaev2006anyons,WitczakKrempa2014, Rau2016, Winter2017, Cao2018, Trebst2017, Schaffer2016}.
Such interactions were proposed to be realizable in the edge-sharing octahedra of $d^5$ transition
metal ions, where strong spin-orbit coupling (SOC) splits the $t_{2g}$ states into multiplets with effective angular momentum $j_{\rm eff}$ = 3/2 and 1/2~\cite{Jackeli2009,chaloupka2013zigzag,rau2014generic,rau2014trigonal}.  While other proposals for realization of the Kitaev model also exist for materials with $d^7$ filling \cite{sano2018kitaev,liu2018pseudospin}, as well as for complex magnetic interactions for $d^1$ filling \cite{yamada2017emergent}, we concentrate in this work on the well-studied $d^5$ case. 
Promising candidate materials include {\nairo}~\cite{Singh2010,Choi2012,Singh2012},
{\liiro}~\cite{Singh2012,Gretarsson2013, modic2014realization,Freund2016},
{\rucl}~\cite{plumb2014alpha,kim2015kitaev, Johnson2015,banerjee2016proximate, banerjee2017neutron}, 
as well as {\hliiro}~\cite{Bette2017,kitagawa2018spin,li2018role}. 

One of the persistent questions regarding all of these spin-orbital coupled magnets is the specific details of the low-symmetry magnetic couplings, which are difficult to extract from any single experiment. The overall scale and anisotropies are often first addressed 
via the (direction-dependent) Weiss constant $\Theta$, appearing in the phenomenological Curie-Weiss (C-W) law describing the high-temperature magnetic susceptibility: 
\begin{equation}
\chi = \chi_0+\frac{N_s(\mu_{\rm eff})^2}{3 k_B (T-\Theta)}, 
\label{eq:curie0}
\end{equation}
where $\chi_0$ accounts for temperature-independent background contributions, $N_s$ is the number of sites, and $\mu_{\rm eff}$ denotes the effective magnetic moment. While thermal fluctuations dominate for temperatures $T \gtrsim \Theta$, quantum effects typically play a decisive role for $T \ll \Theta$. 
For this reason, quantum magnets nearby spin-liquid ground states with finite but suppressed ordering temperature $T_N$ may still display a wide temperature range $T_N < T < \Theta$ where responses resemble those of spin-liquid states. This occurs 
provided a large frustration parameter $f = \Theta/T_N$ can
be defined. The excitations in this temperature regime may even be interpreted in terms of fractionalization~\cite{nasu2016fermionic,do2017majorana,motome2020hunting,hanli2020universal}.
Evidently, accurate estimation of $\Theta$ is an important first characterization 
of a spin-liquid candidate and frustrated magnets in general.
However, as we discuss in this work, standard C-W fits are insufficient for $j_{\rm eff}$ magnets with strong SOC.

For the examples of {\nairo}~\cite{Singh2012} and {\liiro}~\cite{Freund2016}, standard C-W fits suggest strongly anisotropic $\Theta$-values as large as $\sim-125\,$K despite antiferromagnetic ordering temperatures of 15\,K in {\liiro}~\cite{Singh2012, Williams2016} and 13$-$18\,K in {\nairo}~\cite{Singh2012, Ye2012, Liu2011}. 
While competition between anisotropic interactions of different signs may render $f$ a poor measure of frustration \cite{Singh2012}, the scale of the couplings suggested by these $\Theta$-values is much larger than expected from {\it ab-initio} calculations \cite{katukuri2014kitaev,yamaji2014first,Winter2016,das2019magnetic}. 
Furthermore, recent analysis of RIXS data on {\nairo} led to proposed models that account for neither the anisotropy nor the magnitude of the observed Weiss constants \cite{kim2020dynamic}. 

Similarly, the magnetic susceptibility of {\rucl} ($T_N \sim 7\,$K) has been measured by various groups~\cite{Banerjee2017, Kubota2015, Majumder2015, Sears2015, Kim2015,reschke2018sup}, with standard C-W fitting indicating strongly anisotropic Weiss constants up to $\sim 130$\,K, corresponding to $f>15$. This motivated various studies to interpret experimental responses at intermediate temperatures in terms of Kitaev spin-liquid-like behavior \cite{nasu2016fermionic,do2017majorana,hanli2020universal}.
However, theoretical analysis of the inelastic neutron scattering response suggests that the excitation bandwidth may be incompatible with the large energy scales implied by the fitted $\Theta$-values~\cite{winter2017breakdown,laurell2020dynamical}.
Moreover, the fitted effective moments of 2.0 - 2.7\,${\mu}_B$ are anomalously large compared to the pure $j_{\rm eff} = 1/2$ value (1.73\,$\mu_B$), indicating inadequacy of the C-W form. Indeed, similar deviations observed in a wide range of Ru compounds support this conclusion~\cite{lu2018universal}.

The oversimplified use of the Curie-Weiss law can misjudge frustrations~\cite{Nag2017}, relative anisotropies, and signs of the underlying couplings. A key observation is that the Curie-Weiss law only represents an adequate high-temperature approximation for $\chi(T)$ if the quantum operators representing the magnetic moments and magnetic field commute. This holds only if the local moments are of pure spin composition, while strong SOC may induce significant deviations. For isolated paramagnetic metal complexes~\cite{Kotani1949,kamimura1956magnetic,figgis1966magnetic,lu2018universal}, and dimers~\cite{Li2020}, this effect can be modelled by temperature-dependent moments $\mu_{\rm eff}(T)$ due to additional van Vleck contributions. Such effects must also be present in $j_{\rm eff}$ quantum magnets, but are usually ignored in analysis of $\chi(T)$. In this work, we therefore propose an improved formula accounting for $\mu_{\rm eff} (T)$.
We then perform exact diagonalization of the one-site multi-orbital Hubbard model for $d^5$ filling
with inclusion of spin-orbit and crystal-field terms. This allows to compare the results of the improved formula accounting for $\mu_{\rm eff} (T)$ to standard Curie-Weiss fitting for a range of models where the underlying couplings are exactly computed. Finally, we apply the modified fitting formula to  experimental susceptibilities to yield corrected Weiss constants, and discuss  corresponding implications.

\begin{figure}[t]
\includegraphics[angle=0,width=\linewidth]{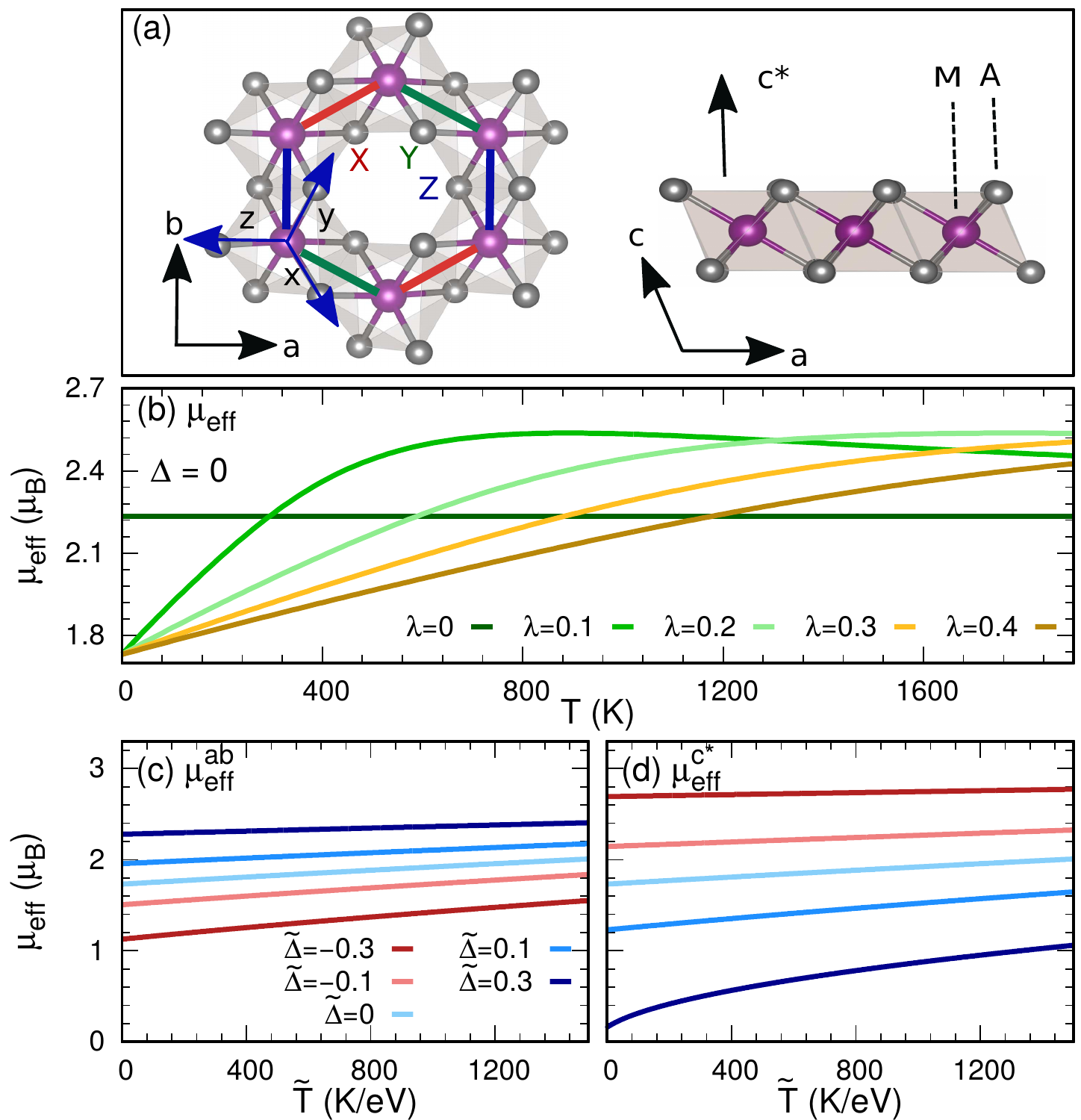}
\caption{(a) Kitaev candidate honeycomb layers ($M=\{\text{Ir},\,\text{Ru}\}$, $A$ = \{O, Cl\}) with view in the $ab$- and $ac$-plane.
	Arrows indicate local $x,\,y,\,z$ directions and crystallographic $a,\,b,\,c$ axes. $c^\ast$ is the direction perpendicular to the $ab$-plane. Canonical X, Y, Z bonds are indicated.
	(b) Temperature-dependent effective magnetic moment $\mu_{\text{eff}}(T)$ for different SOC-strengths $\lambda$ (in eV) and fixed trigonal splitting $\Delta=0\,$eV. 
	(c) $\mu_{\text{eff}}(\tilde{T})$ with $\tilde{T}=T/\lambda$ for different values of
	$\tilde{\Delta}=\Delta/\lambda$ in the $ab$-plane.  (d) $\mu_{\text{eff}}(\tilde{T})$
	perpendicular to the $ab$-plane. }
\label{fig:mu213}
\end{figure}

The electronic Hamiltonian for the $d^5$ filling of octahedrally coordinated 
transition metal ions in edge-sharing geometries (see Fig.~\ref{fig:mu213}(a)) is given by:
\begin{eqnarray}
\mathcal{H}_{\rm tot} = \mathcal{H}_{\rm hop}+\mathcal{H}_{\rm CF}+\mathcal{H}_{\rm SO} + \mathcal{H}_U,
\label{hamil}
\end{eqnarray}
which is the sum of, respectively, the kinetic hopping term, crystal field splitting, spin-orbit coupling, and Coulomb interaction. The explicit expression for each term is given in the Supplemental Material~\cite{Suppl}. Locally, SOC splits the $t_{2g}$ levels into $j_{1/2}$ and $j_{3/2}$ states, with a single hole in the $j_{1/2}$ level in the ground state. The low-energy states are thus spanned by $j_{1/2}$ doublet degrees of freedom, which can be described by an effective spin model with $j_{\rm eff} = 1/2$. 
The effective Hamiltonian is written as $\mathcal{H}_{\rm eff} \equiv \mathbb{P}( \mathcal{H}_{\rm tot} + \mathcal{H}_{\rm Z}) \mathbb{P}$,
where~\footnote{We only consider bilinear magnetic exchange here.}:
\begin{align}
 \mathbb{P} \mathcal{H}_{\rm tot}\mathbb{P} = \sum_{i\alpha j \beta} J_{ij}^{\alpha \beta} S_i^\alpha S_j^\beta, \label{eq:proj} \\
 \mathbb{P} \mathcal{H}_{\rm Z} \mathbb{P} = - \sum_{i \alpha \beta} h_\alpha g_i^{\alpha\beta} S_i^\beta. \label{eq:proj_g}
 \end{align}
Here, $\mathbb{P}$ is a projection operator onto the low-energy subspace, $J_{ij}^{\alpha\beta}$ describe interactions between $j_{1/2}$ pseudospin components $ S_i^\alpha$  
($\alpha \in \{x,y,z\}$),  $g_{i}^{\alpha\beta}$ are effective $g$-values, and $h_\alpha$ respective magnetic field components. 
The conjugate high-energy subspace contains states with finite density of local $j_{1/2} \to j_{3/2}$ spin-orbital excitons, and intersite particle-hole excitations. 
In reality, the Zeeman operator $\mathcal H_\mathrm Z$ mixes the $j_{1/2}$ and $j_{3/2}$ states, generating contributions to the magnetic susceptibility that are not captured within this low-energy theory. Such \textit{van Vleck}-like contributions may modify the high-temperature susceptibility significantly. We therefore consider a regime where the temperature $k_BT$ is large compared to the magnetic interactions between $j_{1/2}$ moments ($k_BT \gg J_{ij}^{\alpha\beta} \sim 10$ -- 100\,K), but small compared to the splitting between the $j_{1/2}$ and $j_{3/2}$ levels ($k_BT \ll \lambda \sim 0.1 - 0.5\,$eV $\sim$ 1160 -- 5800\,K).   
For this case, we propose an improved Curie-Weiss formula for the diagonal components of the susceptibility (details of the derivation 
are given
in \cite{Suppl}):
\begin{align}
\chi^{\alpha}(T) \approx &  \ \chi_0^\alpha  + \ \frac{C^\alpha(T)}{T-\Theta^\alpha},
\label{eq:sus}\\
C^\alpha(T) = & \ \frac{N_s}{3k_B}[\mu_{\rm eff}^\alpha(T)]^2, \label{eq:curie}\\
\Theta^{\alpha} = & \ -\frac{S(S+1)}{3 k_B}\frac{ \sum_{i\gamma j\delta} g_{i}^{\alpha\gamma}J_{ij}^{\gamma\delta}g_j^{\delta\alpha} }{ \sum_{i\gamma}g_i^{\alpha\gamma}g_{i}^{\gamma\alpha}}.
\label{eq:theta0}
\end{align}
In this approximation, the effective temperature dependence of $\Theta^{\alpha}$ is neglected, which is adequate for the present cases (see \cite{Suppl}).  The most important observation is that the temperature dependence of $\mu_{\rm eff}^\alpha (T)$ severely complicates the extraction of $\Theta^{\alpha}$ from experimental susceptibility data.
It is often possible to fit such data to a conventional Curie-Weiss form $\chi=\chi_0 + C/(T-\Theta)$; however, the values of $C$ and $\Theta$ obtained from such fits are not directly relatable to the exchange constants of the low-energy spin model.
The way to proceed in order to extract reliable C-W constants is to first obtain
the effective moment $\mu_{\rm eff}^\alpha (T)$ of a single magnetic site, which can be computed exactly by diagonalizing the local Hamiltonian $\mathcal{H}_{\rm CF} + \mathcal{H}_{\rm SO} + \mathcal{H}_{U}$. For specific cases of trigonal and tetragonal distortions, analytical expressions are also available \cite{kamimura1956magnetic,kotani1949magnetic}. The C-W constants can then be extracted 
by fitting Eq.~\eqref{eq:sus} to the measured $\chi^{\alpha}(T)$.

In what follows we demonstrate this procedure for the case of octahedral transition metal ions with trigonal symmetry, where the $t_{2g}$  electron level is split into an $a_{1g}$ singlet and an  $e_g$ doublet with a splitting equal to $3\Delta = E(a_{1g}) - E(e_g)$. Fig.~\ref{fig:mu213}(b) illustrates
the temperature dependence of $\mu_{\rm eff}(T)$ for $\Delta = 0$ and in
 Figs.~\ref{fig:mu213}(c) and (d) we show $\mu_{\rm eff}(T)$ as a function of $\tilde{T} = T/\lambda$ for different values of $\tilde{\Delta} = \Delta/\lambda$ considering only the $t_{2g}$ orbitals. For Ru$^{3+}$, we take $\lambda = 0.15$\,eV, while for Ir$^{4+}$, we take $\lambda = 0.4$\,eV~\cite{montalti2006handbook}. 
As suggested by these calculations, the effective moment is a generically increasing function of temperature for low-spin $d^5$ compounds, for all orientations of the magnetic field. This implies that $C(T)$ and $\chi(T)$ are anomalously enhanced with increasing temperature entirely due to local van Vleck contributions. As we show next, if such data is fitted with a conventional Curie-Weiss form, it leads to large Curie constants $C > g^2S(S+1)/(3k_B)$ and anomalously antiferromagnetic Weiss temperatures $\Theta$ compared to Eq.~\eqref{eq:theta0}.

\begin{figure}[t]
\includegraphics[angle=0,width=0.98\linewidth]{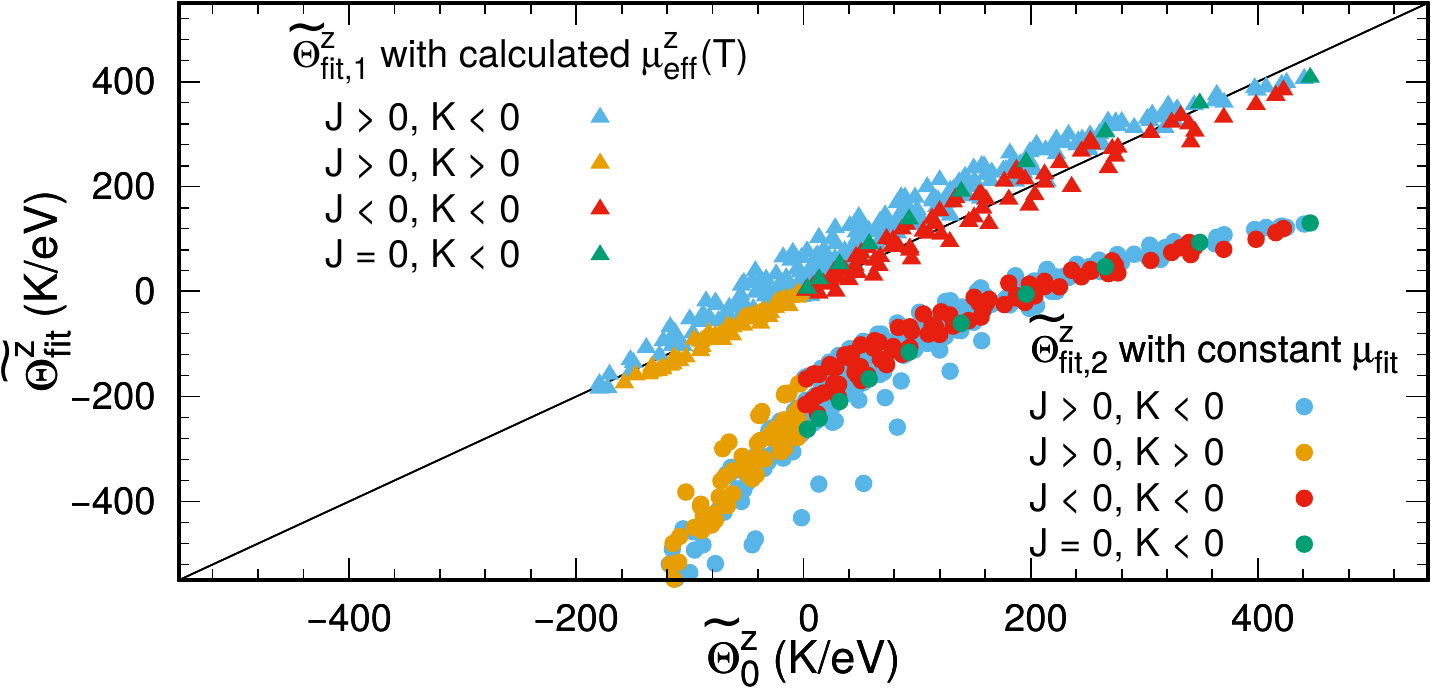} \caption{Comparison
	of fitted renormalized Weiss constant $\tilde{\Theta}_{\text{fit}}$ =
	${\Theta_{\text{fit}}}/\lambda$ for fitting functions
	$\tilde{\Theta}_{\text{fit,1}}^z$
	with temperature-dependent $\mu_{\rm eff}(T)$
	(Eqs.~\ref{eq:sus} and \ref{eq:curie}) and $\tilde{\Theta}_{\text{fit,2}}^z$ with constant $\mu_{\rm fit}$ (Eq.~\ref{eq:curie0})
	vs.\ the intrinsic renormalized Weiss
	constant $\tilde{\Theta}_0$ = ${\Theta_0}/\lambda$ (Eq.~\ref{eq:theta0}), over a wide range of parameters.
The fitted $\tilde{\Theta}_{\text{fit,1}}^z$ agree much better with the intrinsic $\tilde{\Theta}_0$.}
\label{fig:theta2sites}
\end{figure}

In order to benchmark the standard C-W function versus the improved Eq.~\eqref{eq:sus}, we analyze two-site $t_{2g}$-only Hubbard models 
for edge-sharing octahedra with the field oriented perpendicular to the plane of the bond [\textit{i.e.}\ parallel to the cubic $z$-direction, for the canonical Z-bond defined in Fig.~\ref{fig:mu213}(a)]. 
We then consider a range of parameters with $\Delta/\lambda \sim -0.3$ to $+0.3$, $t_2/\lambda \sim 0$ to $1$, and $t_3/\lambda \sim -0.5$ to 0 (see \cite{Suppl} for explicit parameter definitions). 
In the following we discuss results for $U/\lambda \sim 4.25 $ and $J_H/\lambda \sim 0.75 $, corresponding to Ir.
The conclusions below are also valid for parameter values corresponding to Ru.
For each set of hoppings, we first compute the precise low-energy couplings via the projection indicated in Eqs.~(\ref{eq:proj}) and \eqref{eq:proj_g}.
In terms of the cubic ($x,y,z$) coordinates [Fig.~\ref{fig:mu213}(a)], the exchange couplings $J_{ij}$ are conventionally parametrized \cite{rau2014generic,Winter2016} as: 
\begin{align}
J_{ij} = \left(\begin{array}{ccc} J & \Gamma & \Gamma^\prime \\ \Gamma & J & \Gamma^\prime \\ \Gamma^\prime & \Gamma^\prime & J+K \end{array} \right).
\label{eq:interaction_tensor}
\end{align}
From these, we obtain via Eq.~\eqref{eq:theta0} the \textit{intrinsic} Weiss constant $\Theta_0^z = 
[{g_{ab}}^2 (8 {\Gamma'}-2 (\Gamma +3 J+2 K))-4 g_{ab} {g_{c^\ast}} (-\Gamma +{\Gamma'}+K)-{g_{c^\ast}}^2 (2 \Gamma +4 {\Gamma'}+3 J+K)] / [12 {k_B} (2 {g_{ab}}^2+{g_{c^\ast}}^2)]$ where $g_{ab}$ and $g_{c^\ast}$ are the $g$-tensor components in the $ab$-plane and along $c^\ast$. 
We then compute $\chi^z(\tilde{T})$ via full diagonalization of $\mathcal H_{\mathrm{tot}}$ 
(Eq.~\ref{hamil}) on the cluster, and fit it within the region from $\tilde{T} =  800$\,K/eV to 1500\,K/eV, which corresponds to $300 \sim 600$\,K for iridates
and $120 \sim 220$\,K for \rucl.
The results are shown in Fig.~\ref{fig:theta2sites}, where we compare two fitting procedures.
The first fit to $\chi^z(\tilde{T})$, yielding $\tilde{\Theta}_{\rm fit, 1}^z$ ($\Theta_{\rm fit, 1}^z/\lambda$), uses the improved  Eq.~(\ref{eq:sus}) that includes the temperature-dependent $\mu_{\rm eff}(T)$ (determined as described in the previous paragraph). 
The second fit function, yielding $\tilde{\Theta}_{\rm fit, 2}^z$ ($\Theta_{\rm fit, 2}^z/\lambda$), is the standard Curie-Weiss law, with $\mu_{\rm eff}$ 
being a temperature-independent fitting constant. In all cases, we set $\chi_0^\alpha = 0$. 
We find that $\tilde{\Theta}_{\rm fit, 2}^z < \tilde{\Theta}_0^z$ over the entire range of parameters, with deviations from the intrinsic $\Theta_0^z$ as large as $\sim -120$\,K for Ir and $\sim -50$\,K for Ru. In comparison, $\tilde{\Theta}_\mathrm{fit,1}^z$ does not deviate nearly as strong from the intrinsic $\Theta_0^z$.

Having validated the use of Eq.~(\ref{eq:sus}) for a model system, we now turn to the experimental susceptibilities of the $d^5$ Kitaev candidate materials $A_2$IrO$_3$ ($A$ = \{Na, Li\}) and $\alpha$-RuCl$_3$. 
In each case, we make a global fit to data in the $c^*$ axis and $ab$-plane [defined in Fig.~\ref{fig:mu213}(a)] using Eq.~(\ref{eq:sus}) with five fitting parameters: $\chi_0^{c^\ast}, \chi_0^{ab}, \Theta^{c^\ast}, \Theta^{ab}$, and $\Delta$. Note that standard Curie-Weiss fits for these materials employed six free parameters. The effective moments $\mu^\alpha_{\rm eff}(T,\Delta,\lambda)$ were computed via exact diagonalization of $\mathcal{H}_{\rm CF} + \mathcal{H}_{\rm SO} + \mathcal{H}_{U}$ on a single site in each case [as shown previously in Fig.~\ref{fig:mu213}(c) and (d)].
For practical applications, approximative analytical expressions \cite{kamimura1956magnetic,kotani1949magnetic} for $\mu_{\rm eff}$ may be alternatively used.

\begin{figure}[t]
\includegraphics[angle=0,width=\linewidth]{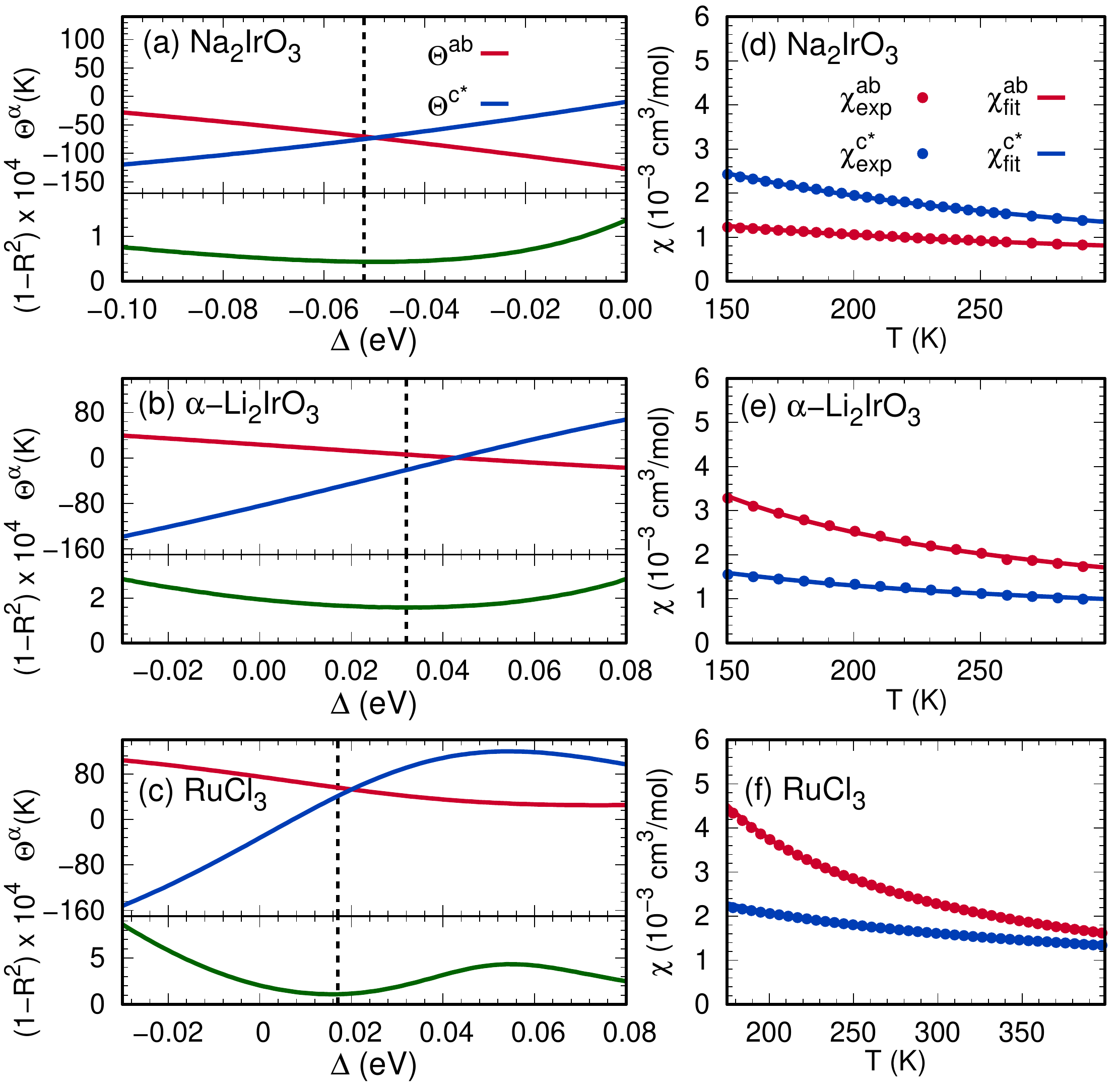}
\caption{
Fits to experimental $\chi(T)$ using Eq.~(\ref{eq:sus}). (a-c): Best fit Weiss constants as a function of crystal field $\Delta$, with $1-R^2$ shown in green indicating fit quality. Vertical dashed lines mark the best overall fits. (d-f): Experimental data from Ref.~\cite{Singh2012,Freund2016,Note2} together with best overall fit.
For each material, $|\chi_0^\alpha|<0.2\cdot10^{-3}\,$cm$^3$/mol, thus influencing the fits negligibly.
The Weiss constants obtained with Eq.~\eqref{eq:sus} differ significantly from conventional Curie-Weiss analysis neglecting the $T$-dependent $\mu_{\rm eff}(T)$. 
\label{fig:theta}}
\end{figure}

The fitting results are presented in Fig.~\ref{fig:theta}. For each compound, we show fitted $\Theta^{c^\ast}$ and $\Theta^{ab}$ as a function of crystal field $\Delta$, together with $(1-R^2)$ to indicate the quality of the fit.  
Below, we discuss the fitted Weiss constants for each compound and their implications for the microscopic couplings by recalling:
\begin{align}
\Theta^{ab}_0 &= -\frac{3}{4k_B}[J + \frac{1}{3} K - \frac{1}{3} (\Gamma + 2\Gamma^{\prime})] \label{eq:thetacal_ab},\\
\Theta^{c^\ast}_0 &= -\frac{3}{4k_B}[J + \frac{1}{3} K  + \frac{2}{3} (\Gamma + 2\Gamma^{\prime})] \label{eq:thetacal_c}.
\end{align}

For Na$_2$IrO$_3$, we refit the susceptibility data from Ref.~\onlinecite{Singh2012} over the range 150 -- 300\,K  [see Figs.~\ref{fig:theta}(a) and (d)].
A standard Curie-Weiss fit yields $\Theta^{ab}_{\rm fit,2} =  -259$\,K and $\Theta^{c^\ast}_{\rm fit,2} =  -90$\,K, which are unlikely to be accurate. Microscopic considerations \cite{rau2014generic,rau2014trigonal, Winter2016} suggest that $\Gamma > 0$, so the finding of $\Theta^{ab}< \Theta^{c^\ast}$ would require very large $\Gamma^\prime <0$, which is broadly incompatible with \textit{ab-initio} calculations \cite{Foyevtsova2013,katukuri2014kitaev,Winter2016} and RIXS experiments \cite{chun2015direct, chaloupka2016magnetic}. 
Using the improved Eq.~(\ref{eq:sus}), we instead obtain $\Theta^{ab}_{\rm fit,1} =  -71\,$K and $\Theta^{c^*}_{\rm fit,1} = -75\,$K. The global best fit corresponds to $3\Delta = -156\,$meV [indicated in Fig.~\ref{fig:theta}(a) by a dashed line], which is compatible with the estimate of $|3\Delta| \sim 170\,$meV from RIXS~\cite{Gretarsson2013}. 
The revised Weiss constants are reduced in magnitude and nearly isotropic, indicating that the anomalous susceptibility anisotropy in this
temperature range likely results from $\mu_{\rm eff}$, \textit{i.e.}\ from the $g$-tensor anisotropy due to the local trigonal distortion.
Assuming the largest nearest-neighbor coupling to be ferromagnetic $K < 0$, the antiferromagnetic sign of the Weiss constants may be explained by further-neighbor antiferromagnetic (Heisenberg) couplings, as previously anticipated for this compound~\cite{kimchi2011kitaev,Winter2016}. Along this line, we note that the revised Weiss constants are in better agreement with a recently proposed model featuring such couplings from Ref.~\onlinecite{kim2020dynamic}
that was inspired by analysis of RIXS measurements (for which $\Theta^{ab}_0 = -73\,$K, $\Theta^{c^*}_0 = -116\,$K). 

Turning to $\alpha$-Li$_2$IrO$_3$, a standard Curie-Weiss fit of the reported susceptibility data \cite{Freund2016} yields $\Theta^{ab}_{\rm fit,2} =  -52\,$K and $\Theta^{c^\ast}_{\rm fit,2} =  -459\,$K. In contrast, for the modified Eq.~(\ref{eq:sus}), the revised Weiss constants are $\Theta^{ab}_{\rm fit,1} = +6\,$K and $\Theta^{c^\ast}_{\rm fit,1} = -21\,$K, which are significantly reduced. The global best fit over a temperature range 150 -- 300\,K corresponds to $3\Delta = +96\,$meV [see Figs.~\ref{fig:theta}(b) and (e)].
Considering Eq.~\eqref{eq:thetacal_ab} and \eqref{eq:thetacal_c}, this relatively small magnitude of the $\Theta$-values may be related to a competition between different couplings, \textit{i.e.}\ a ferromagnetic Kitaev coupling $K <0$, and competitive antiferromagnetic Heisenberg terms (\textit{i.e.}\ $K \sim -3J$). The enhanced anisotropy compared to Na$_2$IrO$_3$ may indicate relatively larger $\Gamma$, $\Gamma^\prime$ couplings. All of these suggestions are consistent with previous {\it ab-initio} estimates \cite{Winter2016}, and place $\alpha$-Li$_2$IrO$_3$ in a region of the $J$-$K$-$\Gamma$-$\Gamma^\prime$ phase diagram \cite{rau2014generic} consistent with the experimentally observed incommensurate ordered state~\cite{williams2016incommensurate}. 

 For \rucl, single crystal susceptibility data from \footnote{Alois Loidl, private communication.}
  is fitted over the temperature range 175 -- 400\,K. A standard Curie-Weiss fit with constant $\mu_{\rm eff}$ yields $\Theta^{ab}_{\rm fit,2} = +35\,$K, $\Theta^{c^\ast}_{\rm fit,2} =-129\,$K, in line with previous reports \cite{Banerjee2017, Kubota2015, Majumder2015, Sears2015, Kim2015}. For the modified Eq.~(\ref{eq:sus}), the global best fit corresponds to $3\Delta = +51\,$meV [see Figs.~\ref{fig:theta}(c), and (f)], which agrees surprisingly well with recent analysis of RIXS data in Ref.~\onlinecite{suzuki2020quantifying}, and Raman scattering and infrared absorption data in Ref.~\onlinecite{warzanowski2020multiple}. For this case, the fitted Weiss constants are $\Theta^{ab}_{\rm fit,1} = +55\,$K and $\Theta^{c^\ast}_{\rm fit,1} = +33\,$K. 
 These values differ significantly in terms of both magnitude and anisotropy from most previous reports (excluding Ref.~\cite{suzuki2020quantifying}). However, they are compatible with the suggested ranges of parameters estimated from \textit{ab-initio} approaches \cite{Winter2016,yadav2016kitaev, kim2016crystal, hou2017unveiling,wang2017theoretical,eichstaedt2019deriving}, employing Eq.~\eqref{eq:thetacal_ab} and \eqref{eq:thetacal_c}.
The overall scale of the couplings also accords with the saturation of nearest-neighbor spin correlations around 
$T \sim \Theta \sim 35\,$K, as measured via 
optical spectral weight for spin-dependent transitions~\cite{sandilands2016optical}. 
 
Assuming that the revised $\Theta$-values are more accurate, we consider their full implications for $\alpha$-RuCl$_3$, as it is the most intensively studied compound. 
For this material, a broad inelastic neutron scattering response reminiscent of the Kitaev spin-liquid ground state was reported for $40\,\mathrm{K}<T < 100$\,K in Ref.~\onlinecite{do2017majorana}. This was discussed in terms of $T_H \sim \Theta \sim 100$\,K, where $T_H$ is an energy scale associated with the Majorana spinon bandwidth.
However, if the true interaction scale is much smaller than these estimates, then this range would instead correspond to the thermal paramagnet ($T > \Theta$), where a relatively wide range of couplings can produce a response similar to the experiment~\cite{Winter2018}. Similarly, in Ref.~\onlinecite{nasu2016fermionic} the temperature dependence of the Raman scattering intensity for $25\,\mathrm K < T <  300$\,K was shown to be compatible with fermionic statistics of the Majorana citations of the Kitaev model. However, the data was modelled with $K \sim 10$ meV, corresponding to $\Theta \sim $ 30 K. Evidently, the majority of the data falls in the thermal paramagnet regime, where coherent magnetic quasiparticles with well-defined statistics are unlikely to persist. 

 In summary, we have investigated the failure of the standard Curie-Weiss law for several Kitaev candidate materials  with strong spin-orbit coupling. For such materials, additional temperature-dependent van Vleck-like contributions always appear, with the lowest-order contribution providing an anisotropic and temperature-dependent effective moment. Failure to account for this effect in fitting of experimental susceptibility yields Weiss constants that are not representative of the underlying magnetic couplings. We therefore proposed and validated a modified 
formula that accounts for $\mu_{\rm eff}(T)$. The latter quantity may be estimated either via exact diagonalization of a local model Hamiltonian, or from analytical expressions \cite{kotani1949magnetic,kamimura1956magnetic} when available. 
This was applied to various $j_{1/2}$ honeycomb materials with $d^5$ filling, and shown to resolve several previous apparent discrepancies between $\chi(T)$ and other experiments. We conclude that some previous reports likely overestimated the scale of the magnetic couplings and possibly the degree of magnetic frustration. For other classes of materials, and other fillings, different deviations may be expected and must be considered. This work should aid in the improved analysis of experimental $\chi(T)$, as a first characterization of novel quantum magnets. 
 
\textit{Acknowledgement}.--- We thank A.\ Loidl, A.\ Tsirlin and P.\ Gegenwart for discussions
and for providing data for \rucl~ and $A_2$IrO$_3$. We also thank
I.I.\ Mazin for useful comments. RV, DAK and
KR acknowledge support by the Deutsche Forschungsgemeinschaft (DFG, German
Research Foundation) for
funding through Project No.\  411289067 (VA117/15-1)
	and TRR 288 --- 422213477 (project A05). 
	YL acknowledges supported by Fundamental Research Funds for the Central Universities (Grant No.\ xxj032019006), China Postdoctoral Science Foundation (Grant No.\ 2019M660249), and National Natural Science Foundation of China (Grant No.\ 12004296).

%


\clearpage
\widetext
\appendix
\begin{center}
\textbf{\large \textit{Supplemental Material}:\\ \smallskip Modified Curie-Weiss Law for $j_{\rm eff}$ Magnets} \bigskip \bigskip
\end{center}
\twocolumngrid

\setcounter{equation}{0}
\setcounter{figure}{0}
\setcounter{table}{0}
\setcounter{page}{1}
\makeatletter
\renewcommand{\theequation}{S\arabic{equation}}
\renewcommand{\thefigure}{S\arabic{figure}}
\renewcommand{\thetable}{S\Roman{table}}

\subsection{Electronic Hamiltonian}

The Coulomb terms of  $H_{tot}$ (Eq. 1 in the
main text) are given by~\cite{Winter2016}:
\begin{align}
\mathcal{H}_{U}& \ = U \sum_{i,a} n_{a,\uparrow}n_{i,a,\downarrow} + (U^\prime - J_{\rm H})\sum_{i,a< b, \sigma}n_{i,a,\sigma}n_{i,b,\sigma} \nonumber \\
 &+ U^\prime\sum_{i,a\neq b}n_{i,a,\uparrow}n_{i,b,\downarrow} - J_{\rm H} \sum_{i,a\neq b} c_{i,a\uparrow}^\dagger c_{i,a\downarrow} c_{i,b\downarrow}^\dagger c_{i,b\uparrow}\nonumber \\ & + J_{\rm H} \sum_{i,a\neq b}c_{i,a\uparrow}^\dagger c_{i,a\downarrow}^\dagger c_{i,b\downarrow}c_{i,b\uparrow},
\end{align}
where $c_{i,a}^\dagger$ creates a hole in orbital $a\in\{yz,xz,xy\}$ at site $i$; $J_{\rm H}$ gives the strength of Hund's coupling, $U$ is the {\it intra}orbital Coulomb repulsion, and $U^\prime=U-2J_{\rm H}$ is the {\it inter}orbital repulsion. The one particle terms are most conveniently written in terms of:
\begin{align}
\vec{\mathbf{c}}_i^\dagger = \left(c^\dagger_{i,yz,\uparrow} \  c^\dagger_{i,yz,\downarrow} \ c^\dagger_{i,xz,\uparrow} \  c^\dagger_{i,xz,\downarrow} \ c^\dagger_{i,xy,\uparrow} \  c^\dagger_{i,xy,\downarrow}\right).
\end{align}
Spin-orbit coupling is described by:
\begin{align}
\mathcal{H}_{\rm SO}=\frac{\lambda}{2} \sum_i \vec{\mathbf{c}}_{i}^\dagger\left(\begin{array}{ccc} 0 & -i \sigma_z & i \sigma_y \\ i \sigma_z & 0 & -i\sigma_x \\ -i \sigma_y & i\sigma_x & 0\end{array} \right)\vec{\mathbf{c}}_i,
\end{align}
where $\sigma_\alpha$, $\alpha=\{x,y,z\}$ are Pauli matrices. The crystal-field Hamiltonian is given by:
\begin{align}
\mathcal{H}_{\rm CF}=  \sum_i \vec{\mathbf{c}}_{i}^\dagger\left\{\mathbf{E}_i\otimes \mathbb{I}_{2\times 2}\right\}\vec{\mathbf{c}}_i,
\end{align}
where $\mathbb{I}_{2\times 2}$ is the $2 \times 2$ identity matrix; the crystal field tensor $\mathbf{E}_i$ is assumed to be:
\begin{align}
\mathbf{E}_i = \left(\begin{array}{ccc} 0& \Delta &\Delta \\ \Delta &0&\Delta \\ \Delta & \Delta & 0 \end{array} \right).
\end{align}
 The hopping Hamiltonian is most generally written:
\begin{align}
\mathcal{H}_{\rm hop}=  \sum_{ij} \vec{\mathbf{c}}_{i}^\dagger \ \left\{\mathbf{T}_{ij} \otimes \mathbb{I}_{2\times 2}\right\}\ \vec{\mathbf{c}}_j,
\end{align}
with the hopping matrices $\mathbf{T}_{ij}$ defined for each bond connecting sites $i,j$. The hopping integrals for the nearest neighbour Z-bond are written as:
\begin{align}
\mathbf{T}_{Z} = \left(\begin{array}{ccc} t_1 & t_2 & t_4 \\ t_2 & t_1 & t_4 \\ t_4 & t_4 & t_3 \end{array} \right).
\end{align}

\subsection{Derivation of Modified Curie-Weiss Law}
For the theoretical derivation of the modified Curie-Weiss law we set the independent background $\chi_0^\alpha=0$. To first determine an expression for the generalized susceptiblity $\chi_\eta$, consider a Hamiltonian $\mathcal{H}=\mathcal{H}_0 + \eta \mathcal{H}_\eta$,  
where $\mathcal{H}_0$ is independent of $\eta$.  The susceptibility of an observable $\mathcal O$ with respect to $\eta$ is defined as $\chi_{\eta} = -\frac{\partial \langle \mathcal O \rangle}{\partial \eta}$, where $\langle \mathcal{O}\rangle = \text{Tr} [ e^{-\beta \mathcal{H}}\mathcal{O} ]/\text{Tr}[e^{-\beta \mathcal{H}}]$ is the thermodynamic expectation value. We assume that $\mathcal O$ itself has no explicit dependence on $\eta$. 
In general, the susceptibility can be computed from:
\begin{align}
\chi_\eta = -\beta \langle \mathcal{H}_\eta \rangle \langle \mathcal{O} \rangle +\int_{-\infty}^{+\infty} d\omega \left(\frac{1-e^{-\beta \omega}}{\omega} \right) \mathcal{C}_{\mathcal{H}_\eta,\mathcal{O}}(\beta,\omega).
\end{align}
$\mathcal{C}_{\mathcal{H}_\eta,\mathcal{O}}(\beta,\omega)$ are temperature-dependent dynamical correlation functions:
\begin{align}\nonumber
\mathcal{C}_{\mathcal{H}_\eta,\mathcal{O}}(\beta,\omega) \equiv \frac{1}{Z}\sum_{n,m}e^{-\beta E_n} \langle n | \mathcal{H}_\eta | m \rangle \langle m | \mathcal{O} | n \rangle  \times \\ \times \delta[\omega - (E_m-E_n)].
\end{align}
where $|n\rangle, |m\rangle$ are eigenstates of $\mathcal{H}$ with energies $E_n, E_m$, and $Z$ is the partition function. 

 In the case where $|n\rangle,|m\rangle$ are also eigenstates of either $\mathcal{O}$ or $\mathcal{H}_\eta$ (i.e.\ $[\mathcal{H},\mathcal{O}] = 0$ and/or $[\mathcal{H},\mathcal{H}_\eta] = 0$), then the correlation function is finite only at $\omega=0$.
 In this case, the susceptibility reduces to:
\begin{align} \label{eqn3}
\chi_\eta = \beta \left( \langle \mathcal{H}_\eta \mathcal{O}\rangle -  \langle \mathcal{H}_\eta\rangle \langle \mathcal{O}\rangle \right).
\end{align}
However, for general operators $\mathcal O$ and $\mathcal H$, this formula does not hold.
Finite-frequency corrections to Eq.~(\ref{eqn3}) include \textit{e.g.}\ van Vleck paramagnetic contributions to the magnetic susceptibility of materials with significant spin-orbit coupling, which we discuss in more detail below.

In general, the Zeeman operator is given by:
\begin{align}
\mathcal{H}_Z = -\mathbf{h} \cdot \mathbf{M} \ \ \ ; \ \ \ \mathbf{M} = \sum_{i} g_s \tilde{\mathbf{S}}_i + g_L \mathbf{L}_i.
\end{align}
Here $\tilde{\mathbf{S}}_i$ denotes the pure spin angular momentum, in contrast to the pseudospin $\mathbf{S}_i$. The magnetic susceptibility tensor is then:
\begin{align}
\chi_h^{\alpha\beta} =-\lim_{h\to 0}\frac{\partial \langle M_\alpha \rangle}{ \partial h_\beta},
\end{align}
where $\alpha,\beta \in \{x,y,z\}$. Since $g_s \neq g_L$, eigenstates of the total angular momentum ($\mathbf{J} = \mathbf{L} + \tilde{\mathbf{S}}_i$) are generally not eigenstates of the Zeeman operator for systems with unquenched orbital angular momentum. It is useful to divide the states into (i) low-energy states, with energies $\omega \approx 0$, which are described by the low-energy spin Hamiltonian, and (ii) high-energy states, with energies $\omega \gtrsim \lambda$. Let $\mathbb{P}$ be the projection operator onto the low-energy space, and let $\mathbb{Q} = 1-\mathbb{P}$ project onto the high-energy space. The dynamical correlation functions can then be divided into two contributions:
\begin{align}
\mathcal{C}_{M_\alpha,M_\beta}(\beta,\omega) = \mathcal{C}_{\alpha,\beta}^{(+)} (\beta,\omega \gtrsim \lambda)+ \mathcal{C}_{\alpha,\beta}^{(-)} (\beta,\omega \approx 0).
\end{align}
The contribution from low-frequency correlations can be computed within the low-energy theory; in the high-temperature limit, it is:
\begin{align}
\int_{-\infty}^{\infty} d\omega \left(\frac{1-e^{-\beta \omega}}{\omega} \right)\mathcal{C}_{\alpha,\beta}^{(-)}(\omega) \approx \beta \text{Tr}[ e^{-\beta \mathcal{H}_{\rm eff}} \mathbb{P} M_{\alpha} \mathbb{P} M_{\beta}\mathbb{P}],
\end{align}
with $\mathbb{P} M_\alpha \mathbb{P} = \sum_{i,\beta} g_i^{\alpha\beta} S_i^{\beta}$ and $\mathcal{H_{\rm eff}}$ as defined in the main text.
Expanding the exponential for large temperatures gives: 
\begin{align}\nonumber 
\lim_{h\to 0} \beta \text{Tr}[ e^{-\beta \mathcal{H}_{\rm eff}} \mathbb{P} M_{\alpha} \mathbb{P} M_{\beta}\mathbb{P}]\approx \\ \nonumber 
 \beta\frac{S(S+1)}{3} \sum_{i,\gamma}g_i^{\alpha\gamma}g_{i}^{\gamma\beta}\\
 -\beta^2 \left(\frac{S(S+1)}{3} \right)^2\sum_{i\gamma j\delta}g_{i}^{\alpha\gamma}J_{ij}^{\gamma\delta}g_j^{\delta\beta} + \mathcal{O}(\beta^3).
\end{align}
If this were the only contribution to the susceptibility, it would be conventional to match these lowest order terms with the expansion of the Curie-Weiss law:
\begin{align}
\frac{C}{T - \Theta} \approx Ck_B \beta + Ck_B^2 \Theta \beta^2 + \mathcal{O}(\beta^3)
\end{align}
from which one would identify:
\begin{align}
C^{\alpha\beta} =  \frac{S(S+1)}{3k_B} \sum_{i,\gamma}g_i^{\alpha\gamma}g_{i}^{\gamma\beta} \label{eq:Curie_27}\\
\Theta^{\alpha \beta} =  -\frac{S(S+1)}{3k_B} \frac{\sum_{i\gamma j\delta}g_{i}^{\alpha\gamma}J_{ij}^{\gamma\delta}g_j^{\delta\beta} }{ \sum_{i,\gamma}g_i^{\alpha\gamma}g_{i}^{\gamma\beta} }. \label{eq:Theta123}
\end{align}
However, this does not account for the high frequency contributions, \textit{i.e.}\ van Vleck-like terms that arise from mixing of the low-energy states with high-energy states:
\begin{align}\nonumber
\mathcal{C}_{\alpha,\beta}^{(+)}(\omega) = 
\frac{\sum_{n,m} e^{-\beta E_n} \langle n| \mathbb{P} M_\mu \mathbb{Q}|m\rangle \langle m | \mathbb{Q} M_\nu \mathbb{P}|n\rangle }{\text{Tr}[\mathbb{P}e^{-\beta \mathcal{H}_0}\mathbb{P}]}  \times \\ \times \delta[\omega - (E_m-E_n)].
\end{align}
This represents two contributions. The first contribution is a temperature-dependent modification of the effective magnetic moment at each site, due to field-induced mixing of different spin-orbital states. The second contribution is a similar modification to the effective intersite interactions. To distinguish these, we further subdivide the high-frequency correlations into single and multi-site correlations:
\begin{align}
\int_{-\infty}^{\infty} d\omega \left(\frac{1-e^{-\beta \omega}}{\omega} \right)\mathcal{C}_{\alpha,\beta}^{(+)}(\omega)& \  \equiv \\ \nonumber \beta\frac{S(S+1)}{3}\sum_i & \Lambda_i^{\alpha,\beta}(\beta) \\ \nonumber - \beta^2 \left(\frac{S(S+1)}{3} \right)^2 & \sum_{ij} \Omega_{ij}^{\alpha,\beta}(\beta) + ...
\end{align}
Here, $\Lambda_i^{\alpha,\beta}(\beta)$ gives the single-site contribution that remains in the limit where all intersite interactions (\textit{e.g.}\ hopping) are taken to zero. In contrast, $\beta^2\Omega_{ij}^{\alpha,\beta}(\beta)$ contains all corrections that result from two-site correlations, \textit{e.g.}\ intersite interactions between excited $j_{3/2}$ and $j_{1/2}$ levels. The factors of $\beta$ are introduced for convenience. 

\begin{figure}[t]
\includegraphics[angle=0,width=\linewidth]{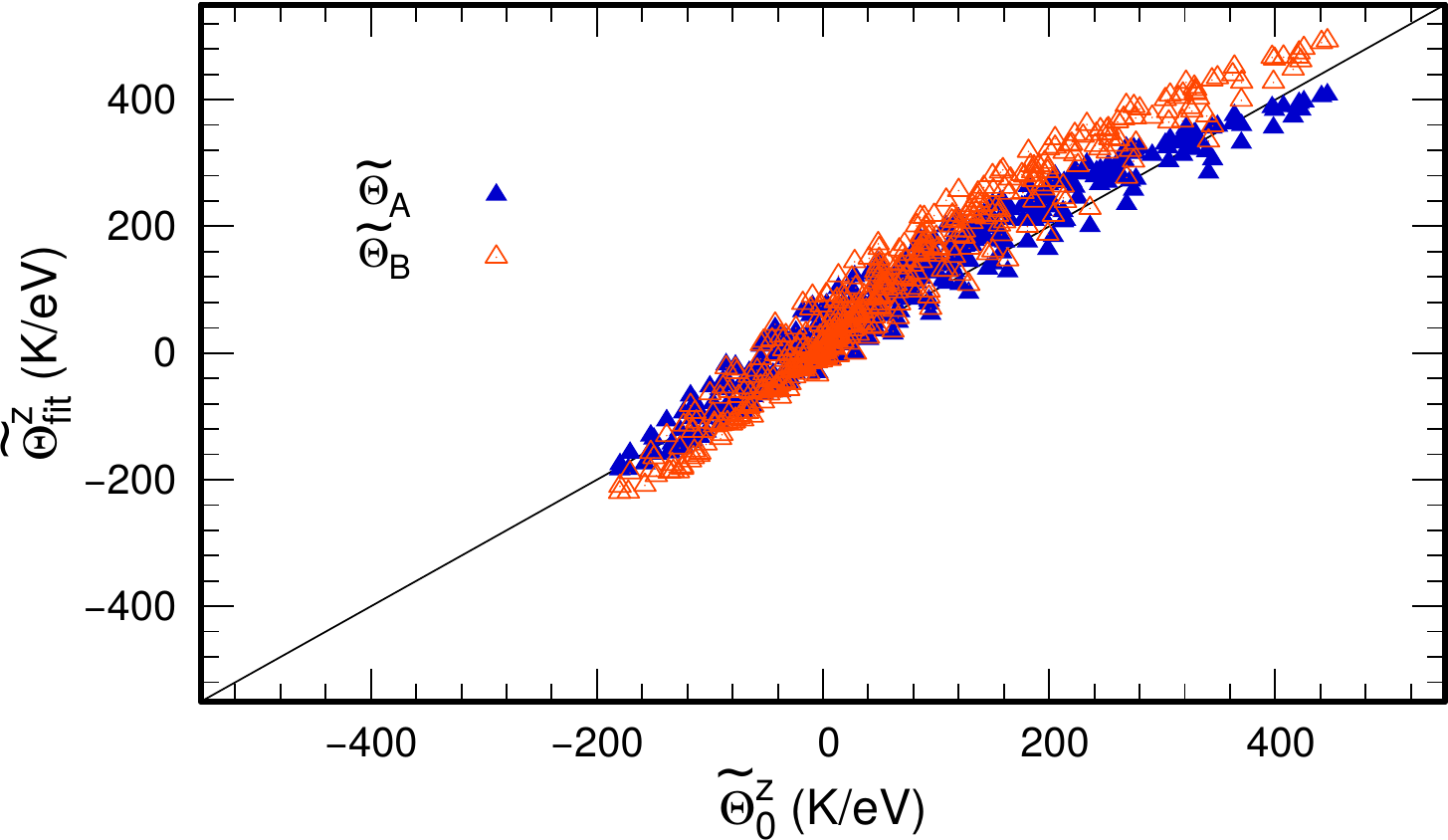}
\caption{Comparison of $\tilde{\Theta}_{\rm fit}^z$ to intrinsic Weiss constant $\tilde{\Theta}_0^z$ for two-site Z-bond model, with blue indicating fits employing the temperature-independent $\tilde{\Theta}_A$ ($\Theta_A/\lambda$), given in Eq.~\eqref{eqn:thetaA}, and orange indicating $\tilde{\Theta}_B$ ($\Theta_B/\lambda$), given in Eq.~\eqref{eqn:thetaB}. The former function performs slightly better.}
\label{fig:thetasup2sites}
\end{figure}

With these contributions included, the Curie and Weiss terms are modified:
\begin{align}
C^{\alpha\beta}(T) =  \frac{S(S+1)}{3k_B} \sum_{i} \left( \Lambda_i^{\alpha\beta}(T) + \sum_{\gamma}g_i^{\alpha\gamma}g_{i}^{\gamma\beta}\right)
\label{eq:Curie_31}
\\
\Theta^{\alpha\beta}(T) =  -\frac{S(S+1)}{3k_B} \frac{\sum_{i j}\left(\Omega_{ij}^{\alpha\beta}(T) +  \sum_{\gamma \delta} g_{i}^{\alpha\gamma}J_{ij}^{\gamma\delta}g_j^{\delta\beta}\right) }{ \sum_{i} \left( \Lambda_i^{\alpha\beta}(T) + \sum_{\gamma}g_i^{\alpha\gamma}g_{i}^{\gamma\beta}\right)  }. \label{eq:Theta456}
\end{align}

Evidently, the van Vleck corrections render both the Curie and Weiss constants temperature dependent, which may complicate the estimation of the low-energy interactions $J_{ij}^{\gamma\delta}$ from temperature-dependent susceptibilities. Restricting now to diagonal susceptibilities (i.e.~$\alpha=\beta$), the modified Curie-Weiss law is written:
\begin{align}\label{eqn:modchi}
\chi_h^{\alpha} = \frac{C^\alpha(T)}{T - \Theta^\alpha(T)} \\
C^{\alpha}(T) = N_s\frac{[\mu^\alpha_{\rm eff}(T)]^2}{3k_B}.
\end{align}
As discussed in the main text, $\mu_{\rm eff}^{\alpha}(T)$ can be estimated for single sites. 
However, $\Omega_{ij}^{\alpha\beta}(T)$ is unknown a-priori, so we have considered two approximations for the Weiss term:
\begin{align}
\Theta_A^\alpha (T)= \Theta_0^{\alpha} =  -\frac{S(S+1)}{3 k_B} \frac{ \sum_{i\gamma j\delta} g_{i}^{\alpha\gamma}J_{ij}^{\gamma\delta}g_j^{\delta\alpha} }{ \sum_{i\gamma}g_i^{\alpha\gamma}g_{i}^{\gamma\alpha}}, \label{eqn:thetaA}\\
\Theta_B^\alpha(T) = \left(\frac{\mu_{\rm eff}(0)}{\mu_{\rm eff}(T)}\right)^2  \Theta_0^\alpha. \label{eqn:thetaB}
\end{align}

In Fig.~\ref{fig:thetasup2sites}, we compare the performance of these approximations for a two-site model of the Z-bond of edge-sharing octahedra with $d^5$ filling and a $j_{\rm eff} = 1/2$ ground state for each site. As in the main text, a range of hoppings was considered. For each set of parameters, the intrinsic $J_{ij}^{\alpha\beta}$ were extracted by numerical projection to the low-energy space, and used to compute the intrinsic $\Theta_0^z$. The susceptibility was then computed exactly, and fit with Eq.~(\ref{eqn:modchi}), using the two approximations for $\Theta^z(T)$. For this case, we find that both approximations yield similar values, with $\Theta_A$ performing better over the parameter range. This suggests that the major deviations are due to the temperature-dependence of the Curie constant, rather than the Weiss constant. 

Note that in Eq.~(7) of the main text and \cref{eq:Theta123,eq:Theta456,eqn:thetaA}, the expressions for $\Theta$ may be significantly simplified for field directions that are principal axes of the $g$-tensor. In particular, if the $g$-tensor is diagonal in the basis, the Weiss constant becomes independent of the $g$-tensor, 
\begin{align} 
\Theta^\alpha = \frac{S(S+1)}{3k_B} \sum_{ij} J_{ij}^{\alpha\alpha}.
\end{align}
In the present case of Kitaev materials, two coordinate system were used: The cubic axes $x$, $y$, $z$ and the crystallographic axes $\hat a=(x+y-2z)/\sqrt{6}$, $\hat b=(y-x)/\sqrt{2}$, $\hat {c^\ast}=(x+y+z)/\sqrt{3}$. The $g$-tensor is approximately diagonal in the $a,b,c^\ast$ basis, while the couplings $J_{ij}$ are usually expressed in the $x,y,z$ basis.

\end{document}